\documentclass[%
reprint,
superscriptaddress,
citeautoscript,
 amsmath,amssymb,
 aps,
 prl,
floatfix,
]{revtex4-1}

\usepackage{graphicx}
\usepackage{dcolumn}
\usepackage{bm}

\begin{document}


\title{Band gap evolution in Ruddlesden-Popper phases}
%

\affiliation{Department of Materials Science and Engineering, University of Delaware, Newark, DE 19716, USA}

\author{Wei Li}
\thanks{These authors contributed equally to this work}
\affiliation{Department of Materials Science and Engineering, University of Delaware, Newark, DE 19716, USA}

\author{Shanyuan Niu}
\thanks{These authors contributed equally to this work}
\affiliation{Mork Family Department of Chemical Engineering and Materials Science, University of Southern California, Los Angeles, CA 90089, USA}


\author{Boyang Zhao}
\affiliation{Mork Family Department of Chemical Engineering and Materials Science, University of Southern California, Los Angeles, CA 90089, USA}

\author{Ralf Haiges}
\affiliation{Loker Hydrocarbon Research Institute and Department of Chemistry, University of Southern California, Los Angeles, California 90089, USA}


\author{Jayakanth Ravichandran}
\email{jayakanr@usc.edu}
\affiliation{Mork Family Department of Chemical Engineering and Materials Science, University of Southern California, Los Angeles, CA 90089, USA}
\affiliation{Ming Hsieh Department of Electrical Engineering, University of Southern California, Los Angeles, CA 90089, USA}

\author{Anderson Janotti}
\email{janotti@udel.edu}
\affiliation{Department of Materials Science and Engineering, University of Delaware, Newark, DE 19716, USA}

\date{\today}
\begin{abstract}


We investigate the variation of the band gap across the Ruddlesden-Popper (RP) series (A$_{n+1}$B$_{n}$X$_{3n+1}$) in model chalcogenide, oxide, and halide materials to understand the factors influencing band gap evolution. In contrast to the oxides and halides, we find the band gap of the chalcogenides evolve differently with the thickness of the perovskite blocks in these natural superlattices. We show that octahedral rotations (\textit{i.e.} deviation of the B-X-B bond angles from 180$^{\circ}$) and quantum confinement effects compete to decide the band gap evolution of RP phases. The insights gained here will allow us to rationally design layered perovskite phases for electronics and optoelectronics.

\end{abstract}

\maketitle

\begin{figure}[!ht]
\begin{center}
\includegraphics[width=4 in]{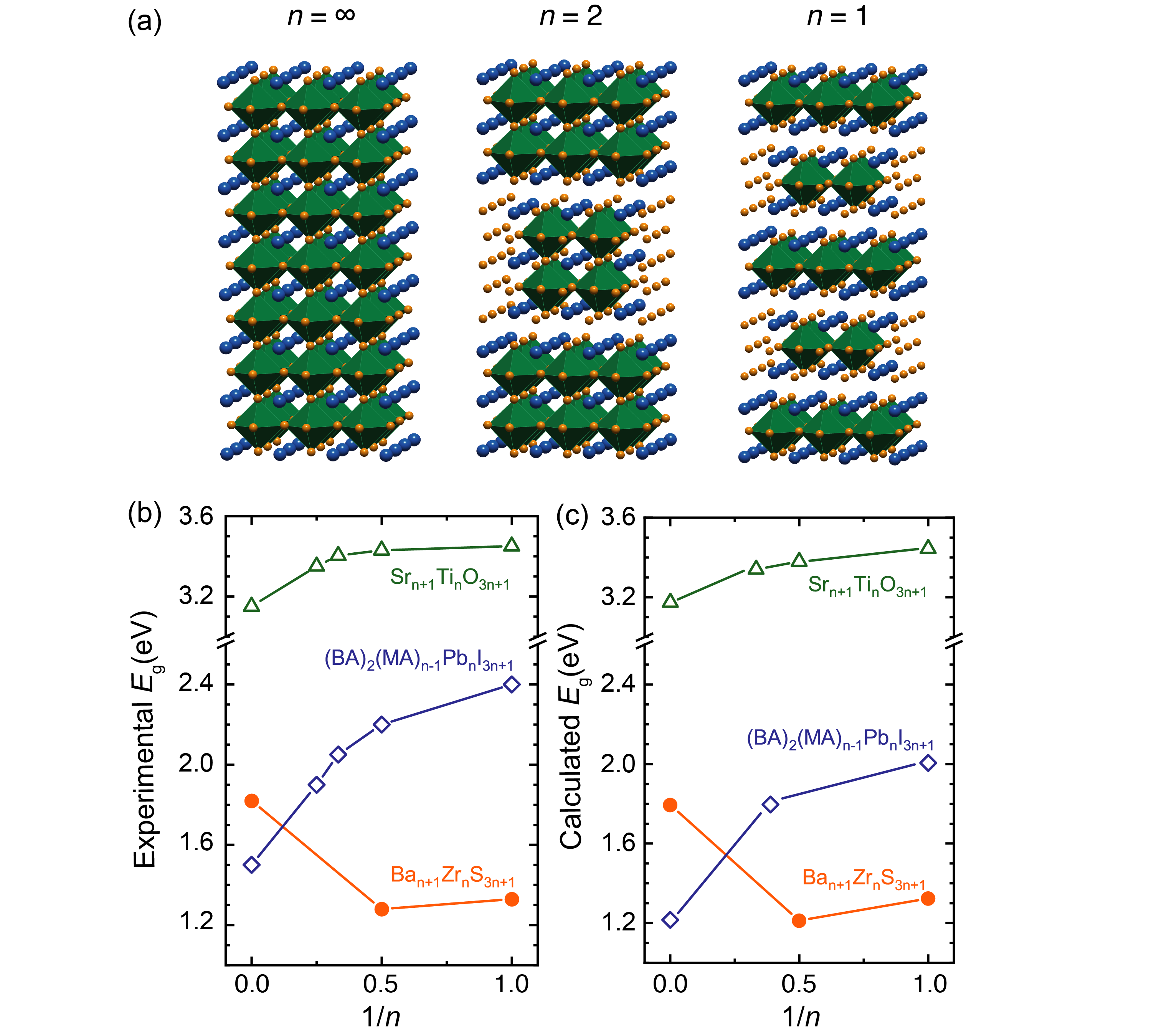}
\end{center}
\caption{(a) Schematic crystal structures of the Ruddlesden-Popper (RP) phases ABX$_{3}$ ($n$=$\infty$), A$_{3}$B$_{2}$X$_{7}$ ($n$=2), and A$_{2}$BX$_{4}$ ($n$=1). Band gaps of Ba-Zr-S, Sr-Ti-O and (BA)(MA)-Pb-I RP series from (b) experiments and (c) calculations. The experimental values of Sr-Ti-O and (BA)(MA)-Pb-I series are from Refs.~\cite{LinzJr1958,Cherian2016,Lee2013,Stoumpos2016,Tsai2016}. The calculated values of the (BA)(MA)-Pb-I band gaps are from Ref.~\cite{Tsai2016,Stoumpos2016}.
}
\label{fig1}
\end{figure}

Perovskites host a variety of emergent phenomena such as  colossal magnetoresistance\cite{Visser1997,Raveau1998,Tokura2006,Murakami2009}, ferroelectricity\cite{Cohen1992,Fong2004,Fan2015}, and superconductivity\cite{He2001,Vaitheeswaran2007}, and offer great structural and chemical flexibility to tune physical properties. The perovskite structure, with a chemical formula of ABX$_3$, consists of an octahedrally coordinated cation (B-site) connected by the corners to form a three dimensional network. Further, perovskites can form two dimensional layered superlattices called as Ruddlesden-Popper (RP) phases, where a perovskite block of varying unit cell thickness $n$ is sandwiched between rock salt layers (AX) to form a natural superlattice. These RP phases possess a general formula of A$_{n+1}$B$_{n}$X$_{3n+1}$ with alternating perovskite blocks displaced by half a unit cell in the in-plane direction [see Fig.~\ref{fig1}(a)]. The perovskite is the end member of the RP series with $n$=$\infty$.

Perovskite oxides containing early transition metals such as Ti, Zr, and main group elements such as Sn are wide band gap semiconductors, but the corresponding chalcogenides possess band gap in the visible-infrared energies. Our ability to
control and modify the band gap of such semiconductors would vastly expand their range of applications, especially as heterostructures.
Alloying is a common approach to tune the band gap of such semiconductors, yet the accompanying disorder has the undesirable effect of lowering carrier mobility\cite{Ogale1984,Rode1983}.
RP phases offer an alternative approach to create long range ordered structures to tune the physical properties such as band gap\cite{Battle1998,Stoumpos2016,Tsai2016,Perera2016,Yu2017,Zhang2017,Chen2018,Li2018}. While the changes in the band gap through alloying can be understood in terms of orbital composition (by chemical substitutions) of the valence and/or conduction-band edge states, it is yet unclear how the quantum confinement through reduced dimensionality, and details of the crystal structure of the RP phases modify the band gap with respect to the ABX$_3$ parent material. A fundamental understanding of the factors at play is crucial to design and tailor emerging perovskite based electronic and photonic materials. 

To motivate our study, we summarize the band gap evolution of three representative materials: Oxide (Sr$_{n+1}$Ti$_{n}$O$_{3n+1}$), halide (BA$_{2}$MA$_{n-1}$Pb$_{n}$I$_{3n+1}$), where MA is CH$_{3}$NH$_{3}$ and BA is CH$_{3}($CH$_{2}$)$_{3}$NH$_{3}$, and chalcogenide (Ba$_{n+1}$Zr$_{n}$S$_{3n+1}$) RP phases in Fig \ref{fig1}(b) and (c)\cite{LinzJr1958,Liu2014,Cherian2016,Lee2013,Stoumpos2016,Tsai2016}. We use a combination of electronic structure calculations and photoluminescence spectroscopy measurements on single crystals of Ba$_{n+1}$Zr$_{n}$S$_{3n+1}$ RP series ($n$=$\infty$,1,2) to determine their band gap and compare them with trends in oxides and halides reported in the literature. We find that the band gap of Ba$_{n+1}$Zr$_{n}$S$_{3n+1}$ are much lower than that of the parent perovskite, BaZrS$_3$,  while the gaps of Sr$_{n+1}$Ti$_{n}$O$_{3n+1}$ and BA$_{2}$MA$_{n-1}$Pb$_{n}$I$_{3n+1}$ are higher than that of SrTiO$_3$ and MAPbI$_3$. 
Our results clearly indicate that the factors controlling the band evolution in chalcogenides are different from oxides and halides. To understand the origin of these contrasting trends, we carried out in depth first principles calculations on the chalcogenide RP phases.

\begin{figure}[t]
\begin{center}
\includegraphics[width=3.2 in]{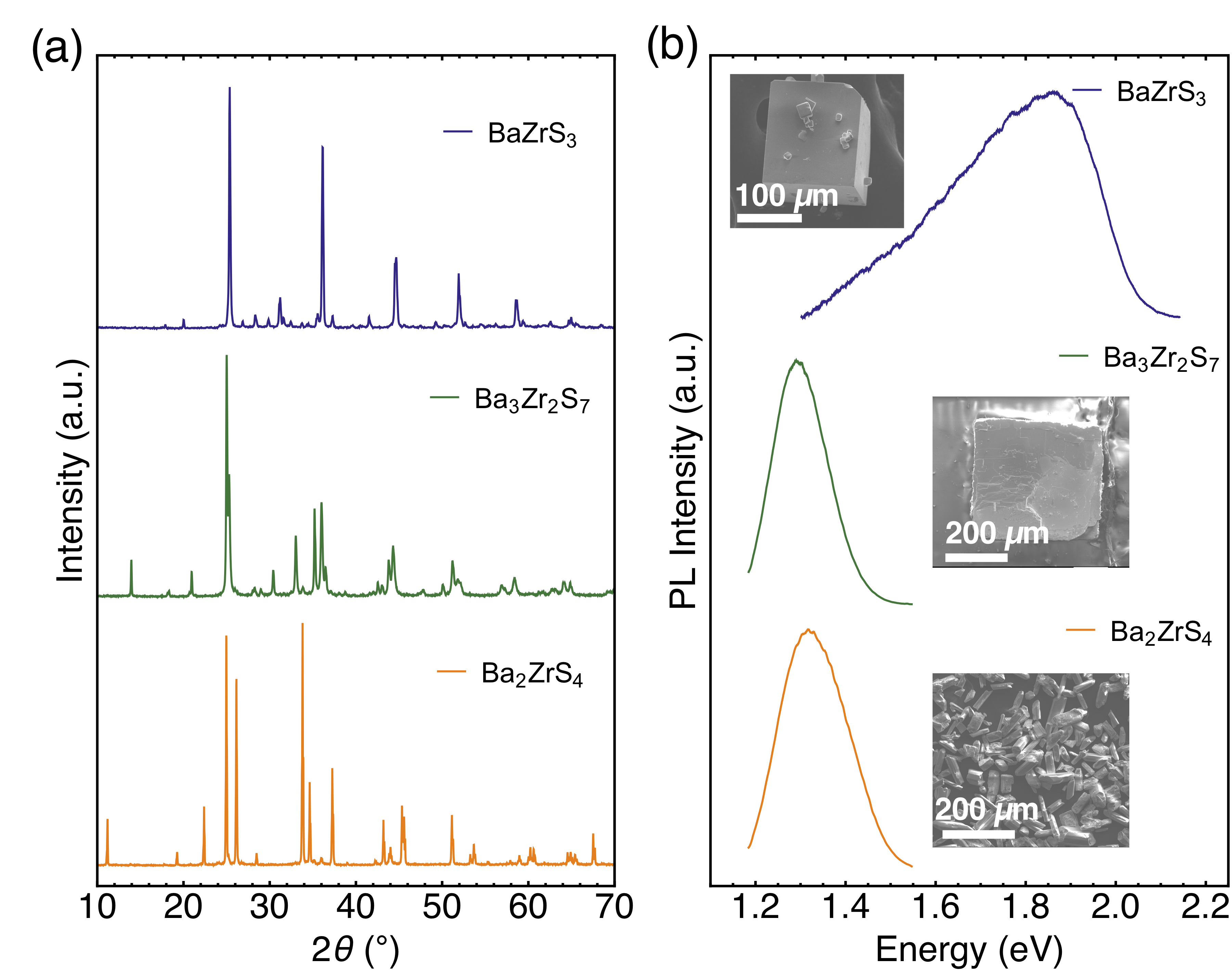}
\end{center}
\caption{Structural and Optical Characterization of the Ba$_{n+1}$Zr$_{n}$S$_{3n+1}$ RP series. (a) X-ray diffraction patterns from ground crystallites of BaZrS$_{3}$, Ba$_{3}$Zr$_{2}$S$_{7}$ and Ba$_{2}$ZrS$_{4}$. (b) Photoluminescence spectra and scanning electron microscopy images (insets) of BaZrS$_{3}$, Ba$_{3}$Zr$_{2}$S$_{7}$ and Ba$_{2}$ZrS$_{4}$ crystallites.
}
\label{fig2}
\end{figure}

The first-principles calculations were performed using density functional theory\cite{Kohn1964,Kohn1965} as implemented in the \textsc{VASP} code\cite{Kresse1993a,Kresse1993b}. The interactions between the valence electrons and the ionic cores are described using projector augmented wave potentials\cite{Blochl1994,Kresse1999}.
For structure optimization, we used the generalized gradient approximation PBEsol\cite{Perdew,Perdew-E} for exchange and correlation, while band structures were computed using the HSE06 hybrid functional\cite{Heyd2003,Heyd2006}. All the calculations were performed with a kinetic energy cutoff of 500 eV for the plane wave basis set. We used $\Gamma$-centered 6$\times$6$\times$6 grid for the 5-atom cubic ABX$_3$ structures, and similar density $k$-meshes for the 20-atom orthorhombic structure, and the tetragonal RP A$_{2}$BX$_4$ and A$_{3}$B$_2$X$_7$ structures. The calculated lattice parameters are listed in the Supplemental Material.

Polycrystalline and single crystal samples of the three Ba-Zr-S RP series compounds were synthesized by solid-state reaction and salt flux growth, respectively, using methods similar to those reported earlier \cite{niu2017,Niu2018}. We performed structural characterization using x-ray diffraction (XRD) and optical characterization using photoluminescence (PL) spectroscopy. We performed single crystal diffraction studies at 100 K. Powder XRD studies were performed on the ground crystallites with Cu K$_\alpha$ radiation in Bragg-Brentano geometry at room temperature. There will be a separate report on the details of crystal growth and X-ray diffraction analysis of the single crystals. Obtained structural parameters are included in the Supplemental Material. PL measurements of the three materials were performed on the single crystals at room temperature with a microscope in back-reflecting geometry.. Emission spectra in 1.5-2.2 eV range were collected in a setup with 532 nm excitation laser while the lower energy range (1.2-1.5 eV) were collected in another setup with 785 nm excitation laser.

\begin{figure}[t!]
\begin{center}
\includegraphics[width=3.2 in]{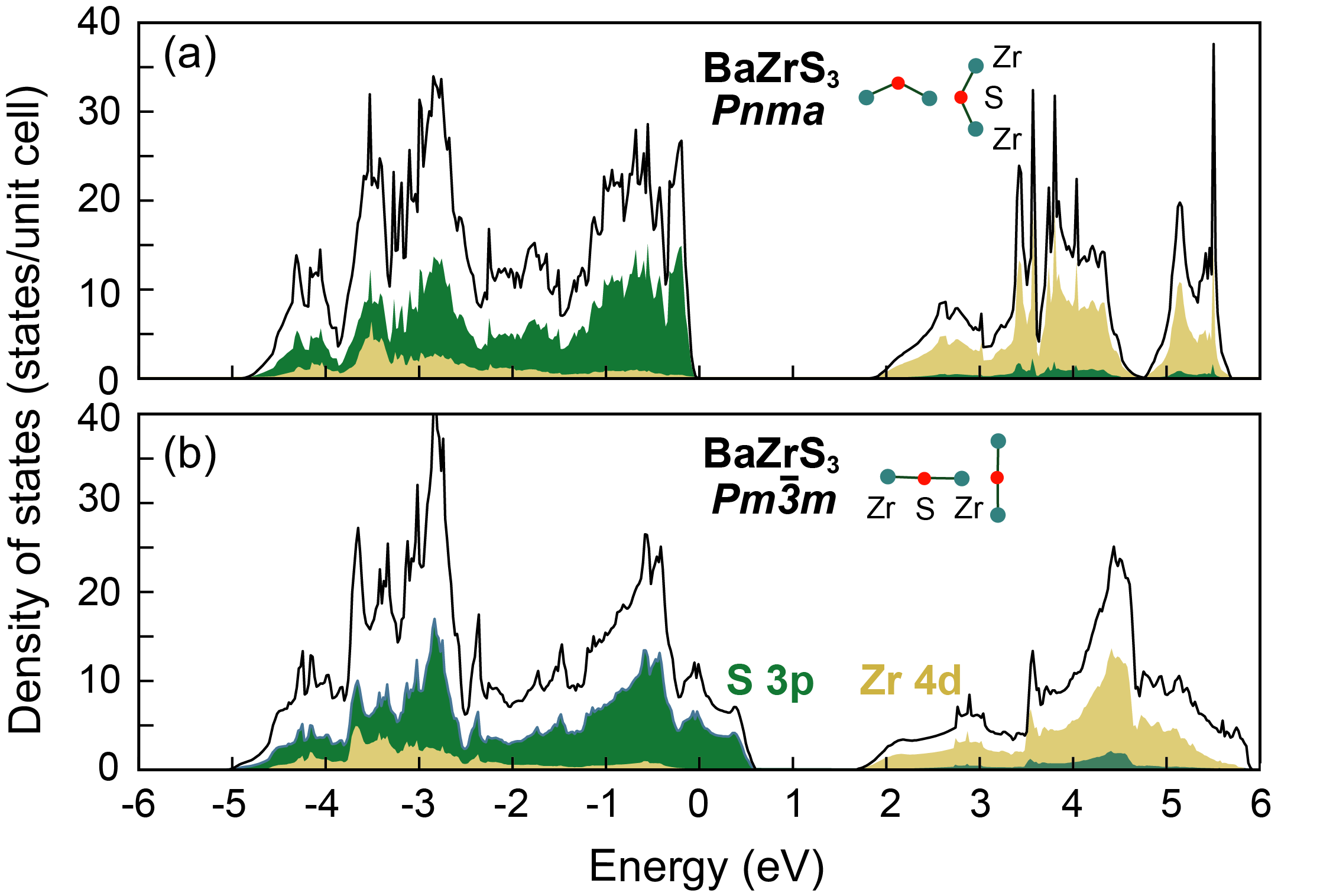}
\end{center}
\caption{Effects of the octahedral rotations on the electronic structure of  BaZrS$_3$. Orbital-projected density of states
of (a) relaxed BaZrS$_3$, featuring octahedral rotations as indicated in the inset, and (b) hypotetial cubic perovskite BaZrS$_3$,
where all the octahedral rotations are removed, keeping the same volume as the relaxed structure. 
The zero in the energy axis is placed at the VBM of relaxed $Pnma$ BaZrS$_3$.
}
\label{fig3}
\end{figure}

As noted earlier, the RP phases of the Sr-Ti-O and (BA,MA)-Pb-I systems follow the trend of increasing band gap with decreasing number of ABX$_{3}$ perovskite layers that are separated by an AX layer. In contrast, we find that for the Ba-Zr-S system, the band gap of BaZrS$_{3}$ is significantly larger than the gaps of Ba$_{3}$Zr$_{2}$S$_{7}$ and Ba$_{2}$ZrS$_{4}$, as shown in Fig.~\ref{fig1}(b)-(c). Our calculated band gaps of BaZrS$_{3}$, Ba$_{3}$Zr$_{2}$S$_{7}$ and Ba$_{2}$ZrS$_{4}$ are 1.79 eV ($n$=$\infty$), 1.21 eV ($n$=2) and 1.33 eV ($n$=1), which are in agreement with past calculations\cite{Li2018,Bennett2009}. To verify this predicted anomalous band gap evolution, we performed PL spectroscopy measurements on single crystals of BaZrS$_{3}$, Ba$_{3}$Zr$_{2}$S$_{7}$ and Ba$_{2}$ZrS$_{4}$. First, we structurally characterized these materials using XRD. The XRD patterns for the ground powders of the single crystals are shown in Fig.~\ref{fig2}(a). The distinct low angle reflections in XRD are clear fingerprints for the RP phases and agree well with the larger unit cells with alternating perovskite and rock-salt layers. The PL spectra and the scanning electron microscopy (SEM) images are shown in Fig.\ref{fig2}(b). BaZrS$_{3}$ ($n$=$\infty$) showed a relatively broad PL peak at around 1.82 eV while the two RP phases showed narrower PL peaks at 1.28 eV ($n$=2) and 1.33 eV ($n$=1) respectively. Thus, our experimental studies confirm the theoretically predicted anomalous band gap evolution in chalcogenide RP phases of Ba-Zr-S.



Now, to address the origin of the anomalous band gap evolution in chalcogenide RP phases compared to the oxides and halides, we need to understand the factors influencing the position and orbital character of the conduction and valence bands in perovskite and RP phases. The valence-band maximum (VBM) and conduction-band minimum (CBM) of the early transition metal (Ti, Zr, Hf) perovskites and related phases have significant contributions from the orbitals of the species that constitute the octahedra (Zr and S in the case of  BaZrS$_3$), and almost none from the orbitals of the A-site species. Specifically, the lowest energy conduction bands are largely composed of transition metal \textit{d} orbitals, while highest energy valence bands are composed of chalcogen \textit{p} orbitals. Therefore, one could expect that the band gap would be significantly influenced by the arrangement of the network of BX$_6$ corner-sharing octahedra, \textit{i.e.} octahedral tilting, rotation, and distortion.
We then discuss how these structural modifications to the highly symmetric corner shared octahedral connectivity affect the band gap of the Ba$_{n+1}$Zr$_n$S$_{3n+1}$ RP series for decreasing thickness of the perovskite blocks, \textit{i.e.} from $n$=$\infty$ to $n$=2 and $n$=1.

First, we compare the calculated orbital-projected density of states of BaZrS$_3$ in the orthorhombic $Pnma$ structure (stable at room temperature) with the hypothetical cubic structure where we removed the octahedral tilt and rotation yet keeping the same equilibrium volume per formula unit. Orthorhombic BaZrS$_3$ displays an in-plane Zr-S-Zr bond angle of 156.6$^\circ$ and out-of-plane Zr-S-Zr bond angle of 160.0$^\circ$, with in-plane Zr-S bond lengths of 2.527 \AA~ and 2.538 \AA, and out-of-plane Zr-S bond lengths of 2.521 \AA, thus featuring a small distortion of the octahedra. By removing the octahedral tilting and rotation the band gap is significantly reduced, by as much as 0.65 eV, as shown in Fig.~\ref{fig3}(a). The octahedral distortion has a negligible contribution to the band gap, changing it by only 0.06 eV. From the orthorhombic to the cubic structure, we find that most of the change in the gap is due to a broadening of the valence band, pushing up the VBM by 0.63 eV, while the CBM is pushed down by only 0.02 eV, as shown in Fig.~\ref{fig3}.  

\begin{figure}[t]
\begin{center}
\includegraphics[width=2.8 in]{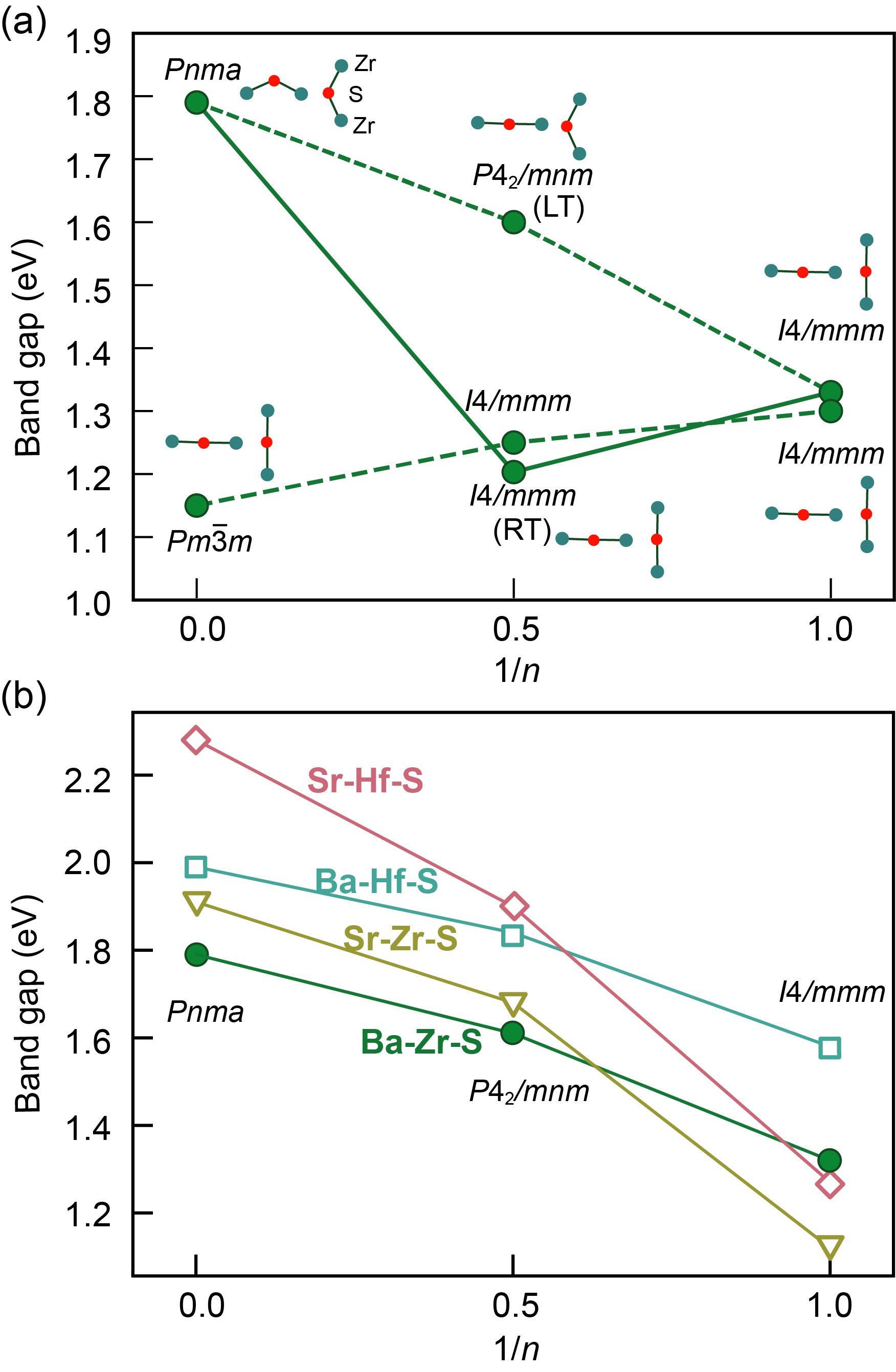}
\end{center}
\caption{Band gap evolution Ruddlesden-Popper phases of in chalcogenides: (a) Calculated band gap of the relaxed BaZrS$_3$, Ba$_{3}$Zr$_{2}$S$_{7}$ and Ba$_{2}$ZrS$_{4}$, compared to the hypothetical phases where the octahedral rotations and Ba/S displacements are removed and the octahedra are made perfect; (b) Calculated band gap of Sr$_{n+1}$Hf$_n$S$_{3n+1}$, Sr$_{n+1}$Zr$_n$S$_{3n+1}$ and Ba$_{n+1}$Hf$_n$S$_{3n+1}$ for $n$=$\infty$, 2, 1 in the $Pnma$, $P4_2/mnm$, and $I4/mmm$ structures, respectively, compared to the Ba$_{n+1}$Zr$_n$S$_{3n+1}$ series.}
\label{Fig4}
\end{figure}


Ba$_{3}$Zr$_{2}$S$_{7}$ ($n$=2) crystallizes in two structures: the low temperature (LT) Ba$_{3}$Zr$_{2}$S$_{7}$ phase \cite{Hung1997}, featuring octahedral rotations, with equilibrium Zr-S-Zr in-plane bond angle of 167.0$^\circ$ and out-of-plane angle of 161.3$^\circ$, and the room-temperature (RT) phase featuring in-plane and out-of-plane angles of 180$^\circ$ yet a small Ba/S displacement \cite{Chen1994}. The calculated band gap for the LT phase is 1.56 eV, and for the RT phase is 1.21 eV. This difference is attributed largely to the lack of octahedral rotations in the RT phase. For example, the removal of all the octahedral rotations and Ba/S displacements(0.45\AA) in the LT phase to make the octahedral arrangements ideal reduces the band gap to 1.25 eV, very close to the result for the RT phase.

On the other hand, Ba$_{2}$ZrS$_{4}$ ($n$=1) compound crystallizes in one stable structure with all the Zr-S-Zr angles at 180$^{\circ}$, \textit{i.e.} there are no octahedral rotations, only octahedral distortion (four in-plane Zr-S bonds of 2.56 \AA~ and two out-of-plane bonds of 2.47 \AA) and a Ba/S (antiferroelectric) displacement(0.35\AA). This high symmetry phase is stable up to the room temperature. The calculated band gap of the relaxed Ba$_{2}$ZrS$_{4}$ is 1.33 eV, compared to 1.19 for a hypothetical structure where the Ba/S displacement is removed, and 1.30 eV for another hypothetical structure, where the Ba/S displacement is removed and the octahedron is made perfect, {\em i.e.}, all Zr-S bond lengths are equal and Zr-S-Zr angles are 180$^{\circ}$. The relatively small differences in the band gap for these three Ba$_{2}$ZrS$_{4}$ structures indicate that the  Ba/S displacement or octahedral distortion have negligible effects on the band gap compared to the octahedral rotations, similar to BaZrS$_3$ ($n$=$\infty$) and Ba$_{3}$Zr$_{2}$S$_{7}$ ($n=2$).

We quantify the effects of octahedral tilting and rotations and quantum confinement on the band gap for the Ba-Zr-S system in Fig.~\ref{Fig4}(a). The lower dashed line represents only the effects of quantum confinement. From ideal cubic BaZrS$_3$ ($n=\infty$, space group $Pm\bar{3}m$ to Ba$_{3}$Zr$_{2}$S$_{7}$ ($n=2$, space group $I4/mmm$) the gap increases by 0.103 eV, and from Ba$_{3}$Zr$_{2}$S$_{7}$ to Ba$_{2}$ZrS$_{4}$ ($n=1$, space group $I4/mmm$), the gap increases by only 0.049 eV. These changes represent purely the effects of quantum confinement, \textit{i.e.} decreasing the thickness of the perovskite blocks from $n=\infty$ to $n=2$ to $n=1$ layer.  This variation in the gap is much smaller than the changes brought about by the octahedral tilting and rotations.  The upper dashed line in Fig.~\ref{Fig4}, connecting relaxed BaZrS$_3$ (space group $Pnma$) to LT Ba$_{3}$Zr$_{2}$S$_{7}$ (space group $P4_{2}/mnm$), and Ba$_{2}$ZrS$_{4}$ (space group $I4/mmm$) contains the effects of octahedral tilting and rotations and the quantum confinement. Since the later is relatively small, as revealed by the lower line, we conclude that the effect of octahedral tilting and rotation on the band gap evolution of the Ba-Zr-S system is dominant. The effects of Ba/S antiferroelectric displacements on the gap of Ba$_{3}$Zr$_{2}$S$_{7}$ and Ba$_{2}$ZrS$_{4}$ are also small, as seen in the differences between the RT Ba$_{3}$Zr$_{2}$S$_{7}$ and ideal Ba$_{3}$Zr$_{2}$S$_{7}$ as well as between the two Ba$_{2}$ZrS$_{4}$ structures, where the upper value corresponds to the experimentally observed phase.

We found that other chalcogenide systems show band gap evolution similar to Ba-Zr-S. As shown in Fig.~\ref{Fig4}(b), the calculated band gap of the Ba-Hf-S, Sr-Zr-S, and Sr-Hf-S RP series also decrease with decreasing $n$.  Our calculations confirm the dominant role of octahedral tilting and rotations in determining band gap, irrespective of the differences in the chemistry. We list all the structural parameters of these materials in Table I and II in the Supplemental Materials. Consistent with our observations for Ba-Zr-S system, the perovskite ABX$_3$ ($n=\infty$) showed greater deviation of the Metal-S-Metal angles from 180$^{\circ}$ compared to the A$_3$B$_2$X$_7$ ($n=2$) and A$_2$BX$_4$ ($n=1$) in all the cases. The deviation quantitatively reflects the degree of octahedral tilting, and proportionately greater band gap values.

In conclusion, our study demonstrates the competition between quantum confinement and octahedral rotations to determine band gap evolution in the layered phases with octahedral coordination such as Ruddlesden Popper (RP) Perovskite phases. We demonstrate that the influence of octahedral rotations is dominant in chalcogenides, unlike oxides and halides, where quantum confinement dictates band gap evolution. We expect that this understanding will be useful to design next generation semiconductors based on such layered phases for electronic and photonic applications.


W.L. and A.J. acknowledge support from the National Science Foundation Faculty Early Career Development Program DMR-1652994. S.N. and J.R. acknowledge the Air Force Office of Scientific Research under award number FA9550-16-1-0335 and Army Research Office under award number W911NF-19-1-0137. S.N. acknowledges Link Foundation Energy Fellowship.
This research was also supported by the the Extreme Science and Engineering Discovery Environment (XSEDE) facility, National Science Foundation grant number ACI-1053575, and the Information Technologies (IT) resources at the University of Delaware, specifically the high performance computing resources.

\bibliography{BaZrS}
\end{document}



\title{Supplemental Material for `Band gap evolution in Ruddlesden-Popper phases'}

\affiliation{Department of Materials Science and Engineering, University of Delaware, Newark, DE 19716, USA}

\author{Wei Li}
\thanks{These authors contributed equally to this work}
\affiliation{Department of Materials Science and Engineering, University of Delaware, Newark, DE 19716, USA}

\author{Shanyuan Niu}
\thanks{These authors contributed equally to this work}
\affiliation{Mork Family Department of Chemical Engineering and Materials Science, University of Southern California, Los Angeles, CA 90089, USA}

\author{Boyang Zhao}
\affiliation{Mork Family Department of Chemical Engineering and Materials Science, University of Southern California, Los Angeles, CA 90089, USA}

\author{Ralf Haiges}
\affiliation{Loker Hydrocarbon Research Institute and Department of Chemistry, University of Southern California, Los Angeles, California 90089, USA}

\author{Jayakanth Ravichandran}
\email{jayakanr@usc.edu}
\affiliation{Mork Family Department of Chemical Engineering and Materials Science, University of Southern California, Los Angeles, CA 90089, USA}
\affiliation{Ming Hsieh Department of Electrical Engineering, University of Southern California, Los Angeles, CA 90089, USA}

\author{Anderson Janotti}
\email{janotti@udel.edu}
\affiliation{Department of Materials Science and Engineering, University of Delaware, Newark, DE 19716, USA}

\date{\today}

\maketitle

\begin{table*}
\caption{Calculated structural parameters for Ruddlesden-Popper Ba$_{n+1}$Zr$_{n}$S$_{3n+1}$ with $n=1, 2, \infty$. Experimental structural parameters obtained from X-ray diffraction studies in this work, as well as previously reported theoretical and experimental structural data are also listed for comparison. Zr-S-Zr bond angles are listed as perpendicular ($\perp$) and parallel ($\parallel$) to the stacking directions. Cal and Exp denote calculated and experimental, respectively. LT, RT, and Ideal denote low temperature, room temperature, and hypothetical structure with Ba/S displacement removed, respectively.}
\centering
\begin{tabular*}{\columnwidth}{ @{\extracolsep{\fill}\hspace{\tabcolsep}} lcccccccclc @{\hspace{\tabcolsep}}}
\hline
\hline
\multirow{2}{*}{Materials} &\multicolumn{3}{c}{lattice parameters (\AA) }& space &\multicolumn{2}{c} {bond angles ($^\circ$)}  & Calculated E$_{g}$ \\
\addlinespace[-2.0ex]
 & $a$ & $b$ & $c$ & group  &\multicolumn{2}{c} {$\perp$ ~~~~~~~ $\parallel$}  & (eV) \\
\hline
BaZrS$_{3}$\\
\hline
Cal (With tilt) &7.075&9.931&6.951&\textit{Pnma}& 156.54 & 159.99  &1.79\\
Cal (No tilt) & 7.016& 7.016 &9.923 & $Pm\bar{3}m$ &180.00 &180.00& 1.15\\
Exp &  7.056 & 9.962 & 6.996 & $Pnma$& 156.91 & 160.24 & \\
Exp\cite{Lelieveld1980} & 7.060 & 9.981 & 7.025 & $Pnma$ & 159.05 & 160.33 & \\
\hline
Ba$_{3}$Zr$_{2}$S$_{7}$ \\
\hline
Cal (LT) & 7.058& 7.058& 25.183 & $P4_{2}/mnm$ &161.29& 167.01& 1.56\\
Cal (RT) & 4.964& 4.964 &25.460 & $I4/mmm$ &180.00 &180.00& 1.21\\
Cal (Ideal) & 4.964& 4.964 &25.460 & $I4/mmm$ &180.00 &180.00&1.25\\
Cal\cite{Zhang2017} & 7.065 & 7.282 & 25.222 & $P4_{2}/mnm$ & 163.38 & 161.55 & \\
Cal\cite{Zhang2017} & 4.970& 4.970 &25.495 & $I4/mmm$ & 180.00 & 180.00 & \\
Exp (100 K) &  7.071 & 7.071 & 25.418 & $P4_{2}/mnm$ & 165.83 & 164.21 & \\ 
Exp (153 K)\cite{Hung1997} &  7.0565 & 7.0565 & 25.371 & $P4_{2}/mnm$ & 165.65 & 164.02 & \\
Exp (295 K)\cite{Chen1994} & 4.9983 & 4.9983 & 25.502 & $I4/mmm$ & 180.00 & 180.00 & \\
\hline
Ba$_{2}$ZrS$_{4}$ \\
\hline
Cal & 4.933& 4.933& 15.657 & $I4/mmm$ &180.00 &180.00& 1.33\\
Cal (Ideal) & 4.933 &4.933& 15.657 & $I4/mmm$ &180.00& 180.00& 1.30 \\
Exp & 4.854& 4.854& 15.911 & $I4/mmm$ & 180.00 &180.00&\\
Exp\cite{SAEKI1991} & 4.785& 4.785& 15.964 & $I4/mmm$ & 180.00 &180.00&\\
\hline
\hline
\end{tabular*}
\end{table*}

\begin{table*}[h!]
\caption{Calculated structural parameters and band gaps of Ba$_{n+1}$Hf$_{n}$S$_{3n+1}$, Sr$_{n+1}$Zr$_{n}$S$_{3n+1}$ and Sr$_{n+1}$Hf$_{n}$S$_{3n+1}$ with $n=1, 2, \infty$. B-X-B bond angles are listed as perpendicular ($\perp$) and parallel ($\parallel$) to stacking directions. Calculated formation energy (eV/f.u.) for distorted ABX$_3$ and A$_3$B$_2$X$_7$ compounds relative to the perfect perovskite phase $Pm\bar{3}m$ and $I4/mmm$ phase are listed.}
\centering
\begin{tabular*}{\columnwidth}{ @{\extracolsep{\fill}\hspace{\tabcolsep}} lcccccccclc @{\hspace{\tabcolsep}}}
\hline
\multirow{2}{*}{Materials} &\multicolumn{3}{c}{lattice parameters (\AA)}& space &\multicolumn{2}{c} {bond angles ($^\circ$)} & E$_{g}$ & $\Delta$H$_{f}$ &$\Delta$E  \\ 
\addlinespace[-2.0ex]
 & $a$ & $b$ & $c$ & group &\multicolumn{2}{c} {$\perp$ ~~~~~~~ $\parallel$} & (eV)&(eV/f.u.)& (eV/f.u.)\\
\hline
SrHfS$_{3}$&7.028&9.653&6.647&\textit{Pnma}&147.13  & 150.47 & 2.28& -9.306&-0.699\\
Sr$_{3}$Hf$_{2}$S$_{7}$&6.968&6.968&23.737&$P4_{2}/mnm$&151.94   & 149.14 & 1.90&-22.469&-0.667\\
Sr$_{2}$HfS$_{4}$&4.850&4.850&14.926&$I4/mmm$&180.00 &180.00 &1.27&-13.032&\\

BaHfS$_{3}$&6.998&9.864&6.932&\textit{Pnma}&158.22 & 161.01 & 1.99&-9.446&-0.146\\
Ba$_{3}$Hf$_{2}$S$_{7}$&6.996&6.996&25.177&$P4_{2}/mnm$&169.26 & 163.75 & 1.74&-23.216&-0.058\\
Ba$_{2}$HfS$_{4}$&4.898&4.898&15.632&$I4/mmm$&180.00  & 180.00 & 1.58&-13.751&\\

SrZrS$_{3}$&6.668&7.096&9.711&\textit{Pnma}& 146.15 & 149.41 & 1.91&-9.225&-0.747\\
Sr$_{3}$Zr$_{2}$S$_{7}$&7.024&7.024&23.732&$P4_{2}/mnm$& 150.33 & 147.37 & 1.68&-22.276&-0.743\\
Sr$_{2}$ZrS$_{4}$&4.887&4.887&14.941&$I4/mmm$& 180.00 & 180.00 & 1.12&-12.868&\\
\hline
\end{tabular*}
\end{table*}

\begin{figure*}[h!]
\begin{center}
\includegraphics[width=3.2 in]{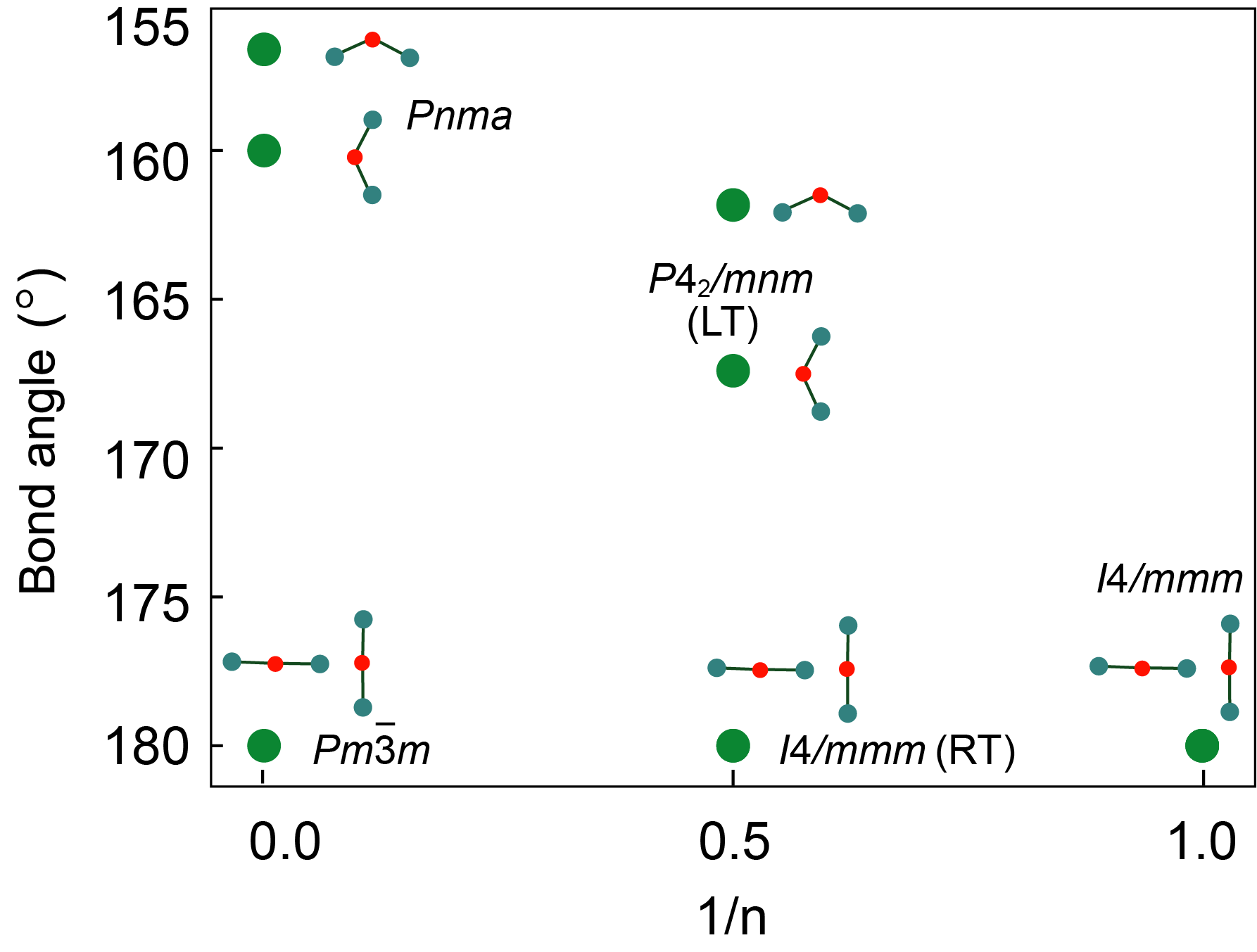}
\end{center}
\caption{Calculated Zr-S-Zr bond angles of the relaxed BaZrS$_3$, Ba$_{3}$Zr$_{2}$S$_{7}$ and Ba$_{2}$ZrS$_{4}$, compared to the ideal phases where the octahedral rotations and Ba/S displacements are removed and the octahedra are made perfect, i.e., with all Zr-S bond lengths are equal.}
\label{fig1s}
\end{figure*}

\newpage

\bibliography{BaZrS}